\documentclass[prb,preprint,showpacs,showkeys,amsfonts,amsmath,amssymb,a4paper]{revtex4}

\usepackage{graphicx}

\begin{document}

\title{In-plane structure and ordering at liquid sodium\\
surfaces and interfaces from \textit{ab initio} molecular dynamics}

\author{Brent G.~Walker}

\affiliation{Industrial Research Limited,69 Gracefield Road, P.~O.~Box
  31-310, Lower Hutt 5040, New Zealand.}

\email{b.walker@irl.cri.nz}

\author{Nicola Marzari}

\affiliation{Department of Materials Science and Engineering,
  Massachusetts Institute of Technology, Cambridge, MA 02139, USA.}

\author{Carla Molteni}

\affiliation{Physics Department, King's College London, Strand,
  London, WC2R 2LS, UK.
}

\begin{abstract}
Atoms at liquid metal surfaces are known to form layers parallel to
the surface. We analyze the two-dimensional arrangement of atoms
within such layers at the surface of liquid sodium, using \textit{ab
initio} molecular dynamics (MD) simulations based on density
functional theory. Nearest neighbor distributions at the surface
indicate mostly 5-fold coordination, though there are noticeable
fractions of 4-fold and 6-fold coordinated atoms. Bond angle
distributions suggest a movement toward the angles corresponding to a
six-fold coordinated hexagonal arrangement of the atoms as the
temperature is decreased towards the solidification point. We
rationalize these results with a distorted hexagonal order at the
surface, showing a mixture of regions of five and six-fold
coordination. The liquid surface results are compared with classical
MD simulations of the liquid surface, with similar effects appearing,
and with \textit{ab initio} MD simulations for a model solid-liquid
interface, where a pronounced shift towards hexagonal ordering is
observed as the temperature is lowered.
\end{abstract}



\thispagestyle{empty}

\maketitle

\thispagestyle{empty}

\section{Introduction\label{sec:intro}}

For a number of years there has been significant interest, from the
theoretical
\cite{rice1981,rice1982,rice1983_2,rice1983_3,rice1984_1,rice1984_2,
rice1987_2,rice1987_3,rice1994,rice1997_2,rice1998_1,rice1998_2,
rice1998_3,rice1999_1,rice1999_2,iarlori,fabricius,chacon1,chacon2,
chacon3,gonzalez,gonzalez2,celestini} and experimental
\cite{rice1978,rice1983_1,expt1,expt2,expt3,expt4,expt5,expt6,expt7,
expt8,expt9,expt10,expt11,rice1987_1,rice1996_1,rice1996_2,rice1997_1,hxray}
points of view, in understanding liquid metal surfaces. This body of
work has established that free liquid metal surfaces exhibit
\textit{surface-induced layering}, in which the liquid atoms near the
surface form into layers parallel to it. This behavior does not
generally occur at free surfaces of liquid dielectrics -- though it is
well known that atomic layers form at the solid-liquid interfaces of
both dielectric and metallic substances \cite{rowlinson-widom} due to
geometrical confinement.  Surface-induced layering at free liquid
metal surfaces was predicted theoretically
\cite{rice1981,rice1982,rice1983_2,rice1983_3,rice1984_1,rice1984_2,rice1987_2,rice1987_3},
and subsequently confirmed experimentally
\cite{expt1,expt2,expt3,expt4,expt5,expt6,expt7,expt8,expt9,expt10,expt11}
for various metals and alloys by x-ray reflectivity measurements. An
understanding of the behavior of liquid metal surfaces is important
for instance to recent studies of studying nanoscale metal alloy
droplets \cite{sutter}.

In addition to providing various hypotheses for the mechanism of layer
formation, simulations have also led to suggestions regarding the
existence of two-dimensional order within the layers. In glue model
simulations of super-cooled liquid Au surfaces, evidence was found by
Celestini \textit{et al} \cite{celestini} for in-plane order with
mostly six-fold coordinated sites, interspersed with five-fold and
seven-fold coordinated ones, in the plane of the surface. They claimed
that a transition to a true hexatic phase \cite{hexatic_note}, in
which each atom is on average six-fold coordinated, is preempted by
solidification of the system. Also of relevance to ordering within the
layers is the experimental work by Reichert \textit{et al}
\cite{five-fold} on liquid Pb in contact with a solid Si substrate;
lead atoms arranged as fragments of icosahedra were observed to be
attached to the substrate, meaning that the Pb atoms were five-fold
coordinated at that interface. A similar atomic arrangement might be
seen near the free surface of a liquid metal if the rapid decay of the
electronic density there acts as a hard wall against which the atoms
pack (as in the suggestion by Rice \textit{et al} for the mechanism of
layer formation
\cite{rice1981,rice1982,rice1983_2,rice1983_3,rice1984_1,rice1984_2,
  rice1987_2,rice1987_3,rice1994,rice1997_2,rice1998_1,rice1998_2,
  rice1998_3,rice1999_1,rice1999_2}, and supported by our simulations
\cite{na_paper_layers,na_paper_jpcmlett}, that makes the analogy with
the formation of atomic layers at solid-liquid interfaces). In
addition, it has been proposed that the natural arrangement of Pb
atoms in the bulk liquid is as icosahedra \cite{spaepen_nature}. So
far however no experimental evidence has been found for in-plane
ordering at free liquid metal surfaces, though investigations using
grazing incidence x-ray techniques are ongoing
\cite{hxray,shpyrko_science}.

We study in detail the properties of free liquid metal surfaces with
extensive \textit{ab initio} molecular dynamics (MD) simulations of
the free liquid surface of sodium. These simulations are made using
the Born-Oppenheimer MD scheme, with the forces acting on the atoms at
each timestep of the dynamics determined via a first principles
electronic structure calculation. The ensemble density functional
theory and the cold-smearing generalized entropy function
\cite{nicolathesis}, with plane wave basis sets and norm-conserving
pseudopotentials, are employed to determine the electronic structure
at each timestep of the MD. This methodology has been demonstrated to
be particularly efficient for studying dynamical properties of
metallic systems \cite{carla_grain_boundary,nicolapaper_al}. Slab
geometries, with periodic boundary conditions, are used to model the
free surfaces, making use of unit cells with two different
cross-sectional shapes parallel to the surface. These two types of
cell, labeled ``(001)'' and ``(111)'', containing 160 and 162 atoms
respectively, are obtained by taking the solid bcc sodium crystal
viewed along either the (001) or (111) directions, melting these as
bulk systems (so there is no vacuum region or surface in each case),
and finally adding large vacuum regions (of $\sim$ 11 \r{A} thickness)
to form liquid slabs, each having two surfaces. Each of the two cell
geometries is simulated at two temperatures ($T=400$~K and $T=500$~K)
above the melting temperature of sodium ($T_{M}=373$~K), meaning a
total of four \textit{ab initio} simulations of the free liquid
surface are undertaken. The use of simulation cells with different
cross-sectional shapes parallel to the surface allows assessment of
whether the cross-sectional geometry has an effect on any in-plane
ordering that might be seen; for instance, it may be that a particular
shape is more or less consistent with a certain type of atomic
arrangement parallel to the surface. Long simulations ($>$50 ps) are
made for each of the two cell shapes at each of the two
temperatures. In each free liquid surface simulation the last 30 ps of
the MD run is used for determination of time-averaged properties.

These dynamical simulations provide detailed microscopic information
from which various properties related to the arrangements of the atoms
at the surface can be extracted. Full details of the MD simulations
are laid out in Refs.~\onlinecite{na_paper_layers} and
\onlinecite{na_paper_jpcmlett}, where we concentrate on understanding
the mechanism of layer formation, in particular with respect to the
relevance of Friedel oscillations and geometrical confinement effects
at free liquid metal surfaces, and the spacing and structure of the
layers normal to the surface.

In the present paper we focus on the two-dimensional atomic
arrangements \textit{within} the layers formed at the liquid sodium
surface. We complement the \textit{ab initio} MD simulations of the
free liquid surface with classical MD simulations of the free liquid
surface, and \textit{ab initio} MD simulations of a model of the
solid-liquid interface.

\section{Results and Discussion \label{sec:results_section}}

\subsection{\textit{Ab initio} MD liquid surface simulations \label{subsec:ab_initio}}

The atomic arrangements parallel to the surface are considered firstly
by looking at density profiles along the surface normal. These exhibit
clear oscillations at the surface, indicating surface-induced
layering. In all our \textit{ab initio} free surface simulations
(where the systems contain $\sim$160 atoms), 7 layers are formed
\cite{na_paper_layers}. While the presence of layers throughout the
slabs is due to their limited sizes, we have demonstrated
\cite{na_paper_layers} that the layered structures of the two surfaces
in each case are independent of one other.

\begin{figure}[!hbt]
\begin{center}
\includegraphics[clip,width=0.9\columnwidth]{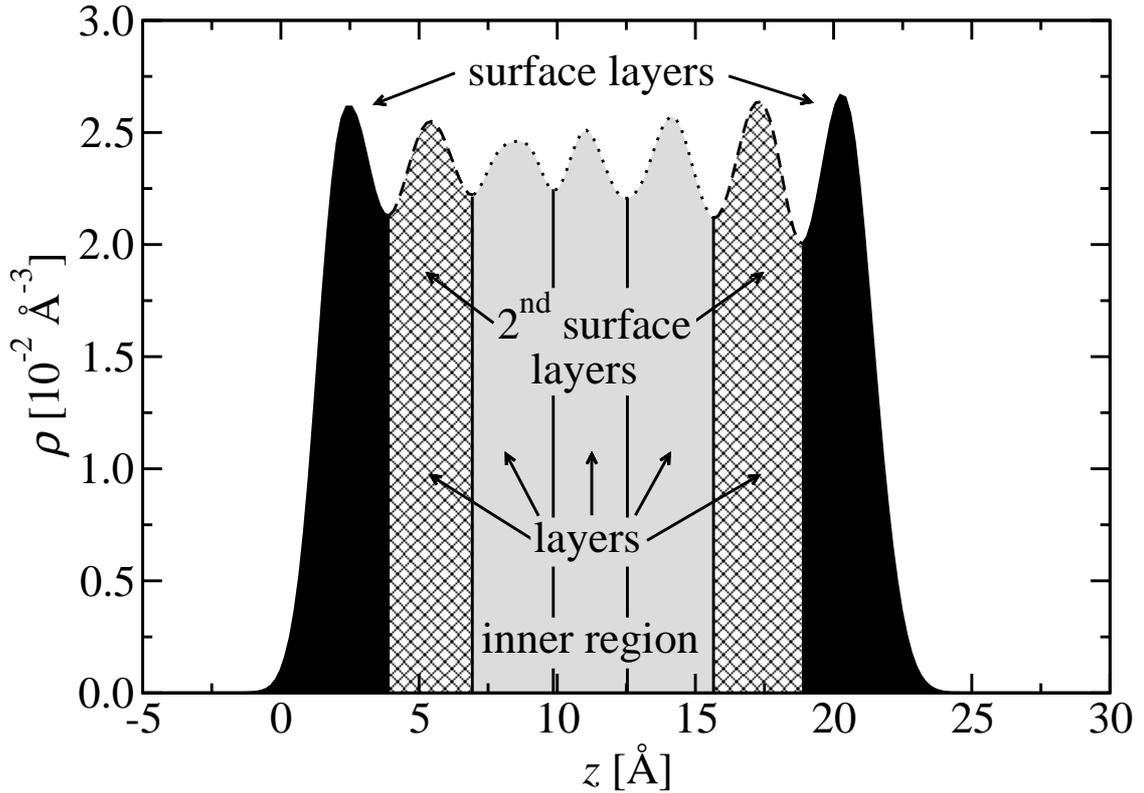}
\caption{\label{fig:interface_pic2} Density profile through slab
  [(001) cell, $T=400$ K] illustrating the regions considered as
  surface layers (black), 2nd surface layers (hashed) and inner (grey)
  for calculation of pair correlation functions, and bond angle and
  nearest neighbor distributions.}
\end{center}
\end{figure}

To examine the atomic structure within the surface layers -- our
primary concern here -- we determine a number of properties: pair
correlation functions, nearest neighbor distributions, bond angle
distributions and two-dimensional density plots in planes parallel to
the surface. The parts of the slabs considered as the surface layers,
the 2nd surface layers and the inner regions for calculating these
various properties are illustrated in Fig.~\ref{fig:interface_pic2} by
way of one of the transverse density profiles (that obtained for the
(001) cell) at $T=400$ K): we emphasize that we take the surface
layers to be those regions on either side of the slab beyond the first
minimum in the density profile.

\begin{figure}[!hbt]
\begin{center}
\includegraphics[clip,width=\columnwidth]{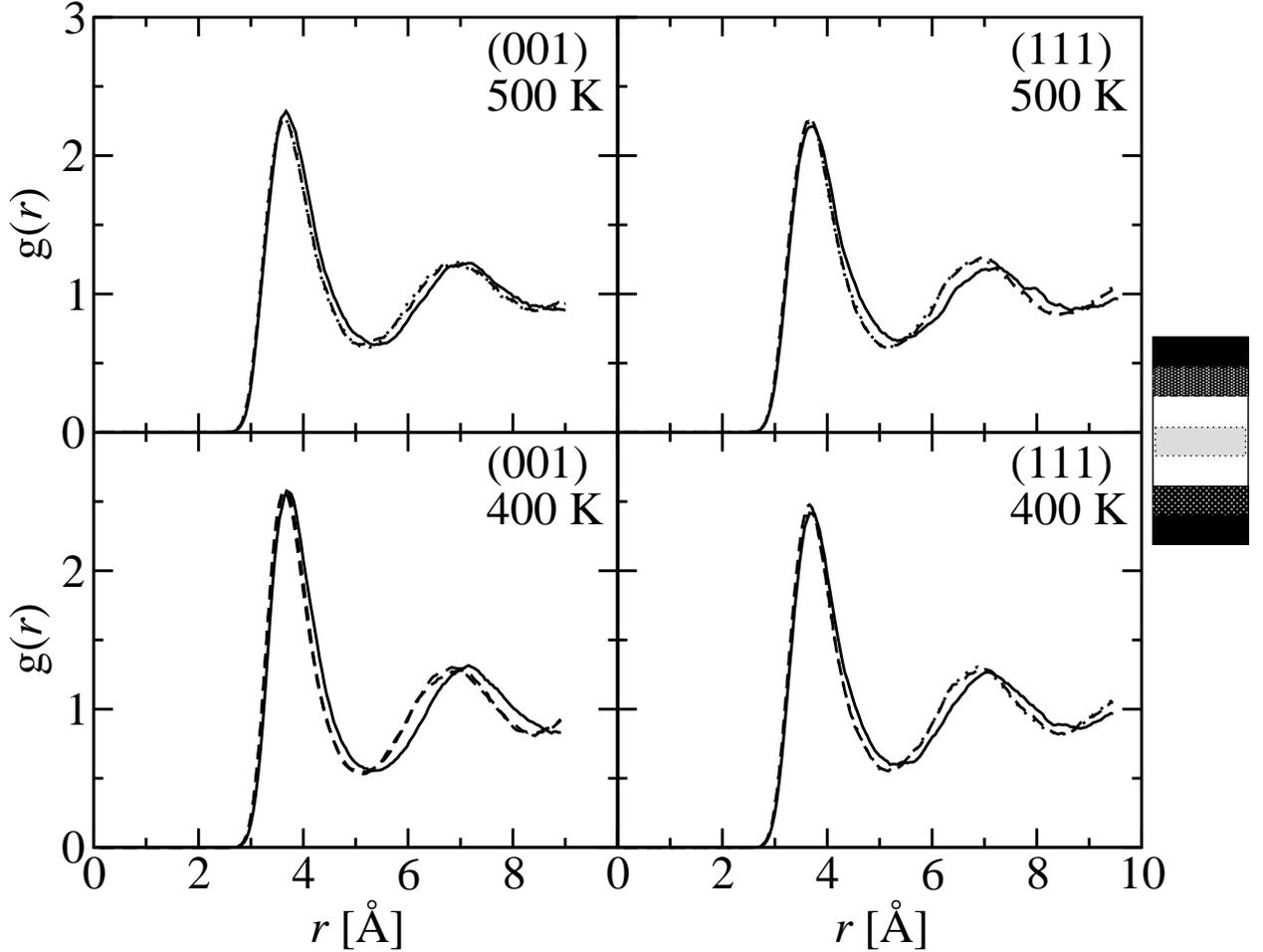}
\caption{\label{fig:corr_slab_pic} Transverse atomic pair correlation
  functions for sodium from \textit{ab initio} MD surface
  simulations. Solid lines: averaged over surface layers; dashed
  lines: averaged over 2nd surface layers; dotted lines: averaged over
  layers in the inner regions. The bar at the right indicates the
  regions of the simulation slabs considered in each case.}
\end{center}
\end{figure}

Transverse atomic pair correlation functions
\cite{rice1987_3,rice1998_1,rice1998_2,rice1998_3,rice1999_2} for the
surface layers, the 2nd surface layers, and slices within the bulk
regions appear in Fig.~\ref{fig:corr_slab_pic}. In comparing the
correlation functions in the surface layers with those for the 2nd
surface layers and the inner parts of the slabs (obtained by averaging
over \textit{all} the inner layers), we see that the positions of the
first peaks are all essentially the same. In all cases however, there
is an observable shift of the second peak to larger values (of
$\sim$3$\%$) at the surface. The 2nd surface layers and the inner
regions have coincident pair correlation functions. This shift in the
second peak in the correlation functions at the surface indicates that
the \textit{second} coordination shell is less tightly packed at the
surface than in the 2nd surface layers and the inner regions. The
shifts observed in the peaks corresponding to the second coordination
shell are consistent with the behavior observed in the orbital free
DFT results of Ref.~\onlinecite{gonzalez3}; however we note that in
Ref.~\onlinecite{gonzalez3} shifts toward larger radii are also seen
in the peaks corresponding to the \textit{first} coordination shells
for Na and Li, that we do not observe in our \textit{ab initio} free
surface simulations. The most probable nearest neighbor distance,
calculated as the average position of the first peak in the pair
correlation function is $r_{\mathrm{nn}}=3.8$~\r{A}.

\begin{figure}[!hbt]
\begin{center}
\includegraphics[clip,width=0.9\columnwidth]{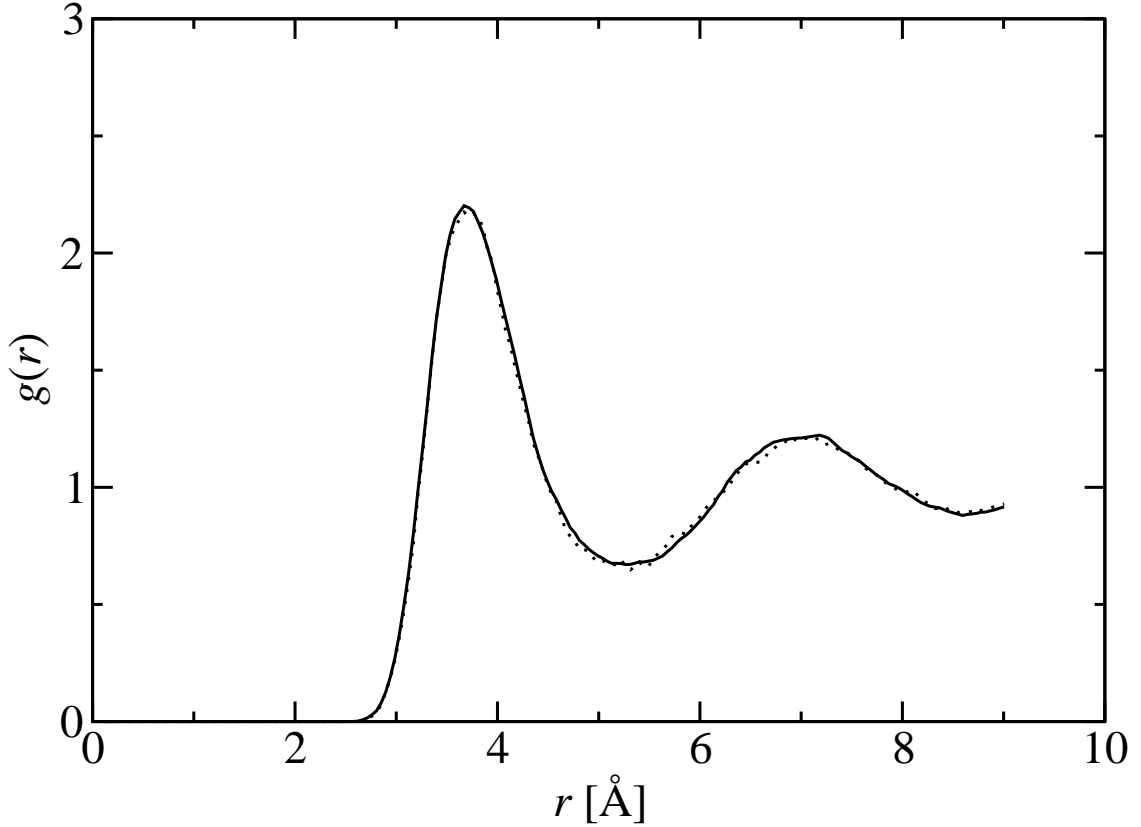}
\caption{\label{fig:corr_bulk_pic} Comparison between the
  three-dimensional pair correlation function calculated in the inner
  region of the slabs (solid curve) and the average transverse pair
  correlation function for slices of comparable thickness to the
  surface regions, calculated within the inner region (dotted
  curve). The data refers to the $(001)$ simulation cell at
  $T=500$~K.}
\end{center}
\end{figure}

For comparison, the transverse pair correlation functions averaged
over a number of slices of similar thickness to the surface regions
are compared with the three-dimensional bulk pair correlation function
for all of the inner region in Fig.~\ref{fig:corr_bulk_pic}. It is
clear that they coincide, meaning that the differences in the
transverse correlation function at the surface and in the inner
regions are genuinely due to differences in atomic arrangements at the
surface.

\begin{figure}[!htb]
\begin{center}
\includegraphics[clip,width=\columnwidth]{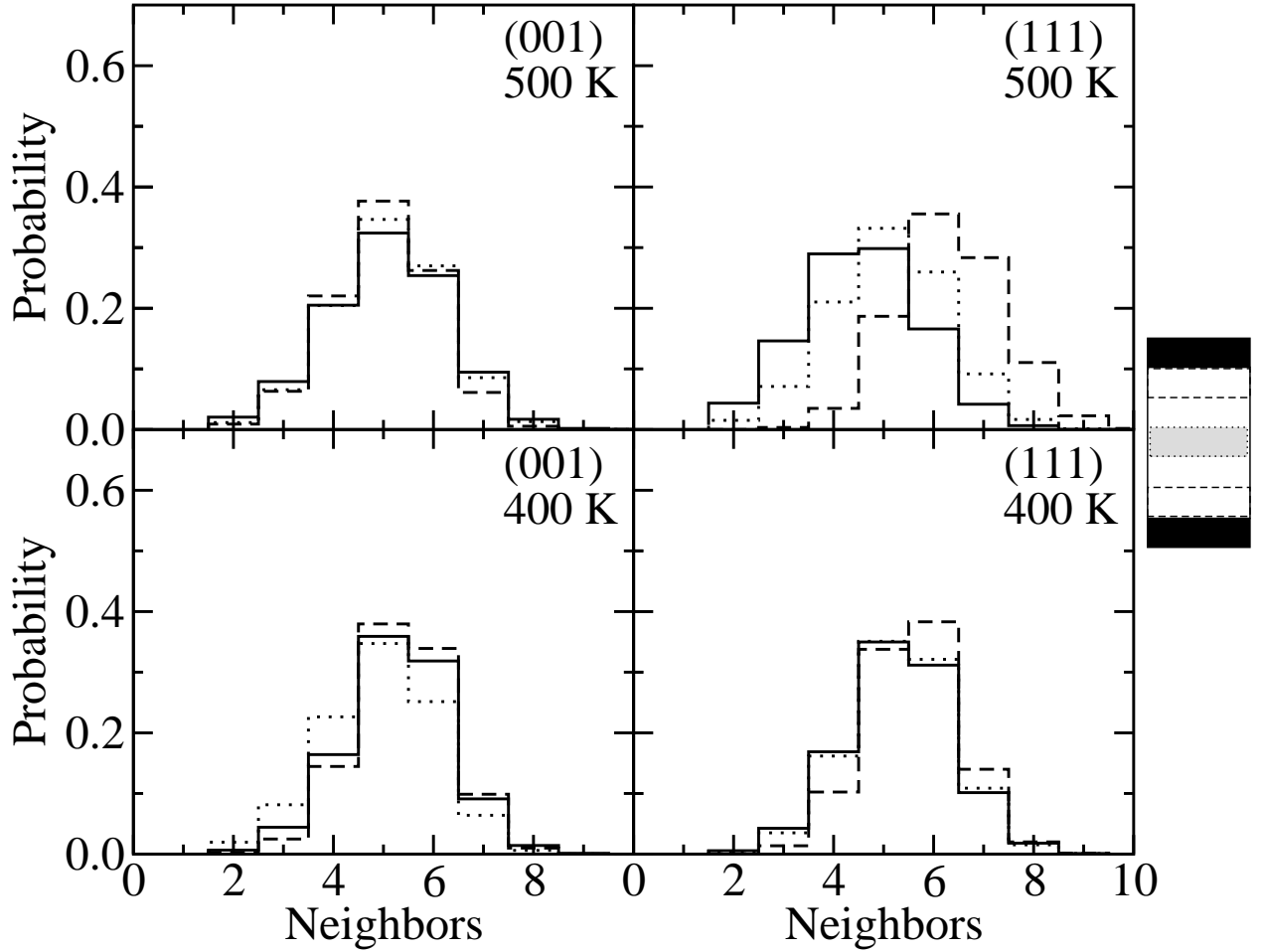}
\caption{\label{fig:nearest_neighbours_slab} Nearest neighbor
  distributions from \textit{ab initio} slab simulations. Solid
  profiles: surface layers; dashed profiles: 2nd surface layers;
  dotted profiles: layers in inner region.}
\end{center}
\end{figure}

In Fig.~\ref{fig:nearest_neighbours_slab} we show the nearest neighbor
distributions obtained from the surface simulations. These were
calculated by counting for each atom the number of atoms within a
sphere of a chosen radius centered on that atom; for the radius of
this ``nearest neighbor'' sphere, we used the value
$r_{\mathrm{cut}}=5.32$~\r{A}, which is the position of the first
minimum in the bulk liquid pair correlation function. Reasonable
variations in the value used for the radius of the nearest neighbor
sphere do not in fact observably affect the nearest neighbor
distributions (the value used is a sensible choice, lying roughly
midway between the first minima in the surface and inner transverse
pair correlation functions shown in Fig.~\ref{fig:corr_slab_pic}).

The majority of atoms in the surface layer have 5 neighbors, though
there are significant fractions ($\sim$30$\%$) with coordination
numbers of 4 and 6 (the latter accounting for the higher fraction of
the two). When the temperature is lowered, the percentage of 4-fold
coordinated atoms decreases, and the percentages of atoms 5 and 6-fold
coordinated increase. This is consistent with the presence of
\textit{icosahedral} fragments, in which each atom is 5-fold
coordinated (as seen in the experimental studies of liquid Pb in
contact with solid Si \cite{five-fold}, and in line with the suggested
arrangement of atoms in the bulk liquid metals as icosahedra
\cite{spaepen_nature}), mixed with 6-fold coordinated regions (as
suggested by Celestini \textit{et al} \cite{celestini}), the latter
regions becoming larger as the temperature is lowered and
correspondingly the disorder is reduced. Coordination numbers obtained
from the nearest neighbor distributions are shown in Table
\ref{tab:nearest_neighbours_table} and are generally just over 5 for
the surface regions of Fig.~\ref{fig:interface_pic2}.

\begin{table}[!htb]
\begin{center}
\begin{tabular}{ccccccc}
\hline $T$ [K] & \multicolumn{2}{c}{Surface Layer} & \multicolumn{2}{c}{2nd Surface Layer} & \multicolumn{2}{c}{Inner} \\
               & (001) & (111) & (001) & (111) & (001) & (111) \\
\hline 400     &  5.3  &  5.3  &  5.5  &  5.6  &  4.9  &  5.3  \\
       500     &  5.1  &  4.5  &  5.0  &  6.3  &  5.1  &  5.1  \\
\hline
\end{tabular}
\end{center}
\caption{\label{tab:nearest_neighbours_table} Coordination numbers
  $n_{c}$ obtained by averaging nearest neighbor distributions shown
  in Fig.~\ref{fig:nearest_neighbours_slab}.}
\end{table}

These average coordination numbers are significantly less than the
value of $n_{c}^{\mathrm{bulk}}=13.14$ obtained from a 500 K bulk
liquid simulation we performed using the same DFT MD methodology. This
is around the value of 13 typical of a bulk liquid, and in agreement
with the value of 13 obtained for Na in earlier bulk sodium
calculations \cite{kresse-hafner}.

We note that the (111), $T=500$ K density profile is less than ideally
symmetric, and this has something of an adverse effect on the other
properties determined for that simulation.

\begin{figure}[!hbt]
\begin{center}
\includegraphics[clip,width=\columnwidth]{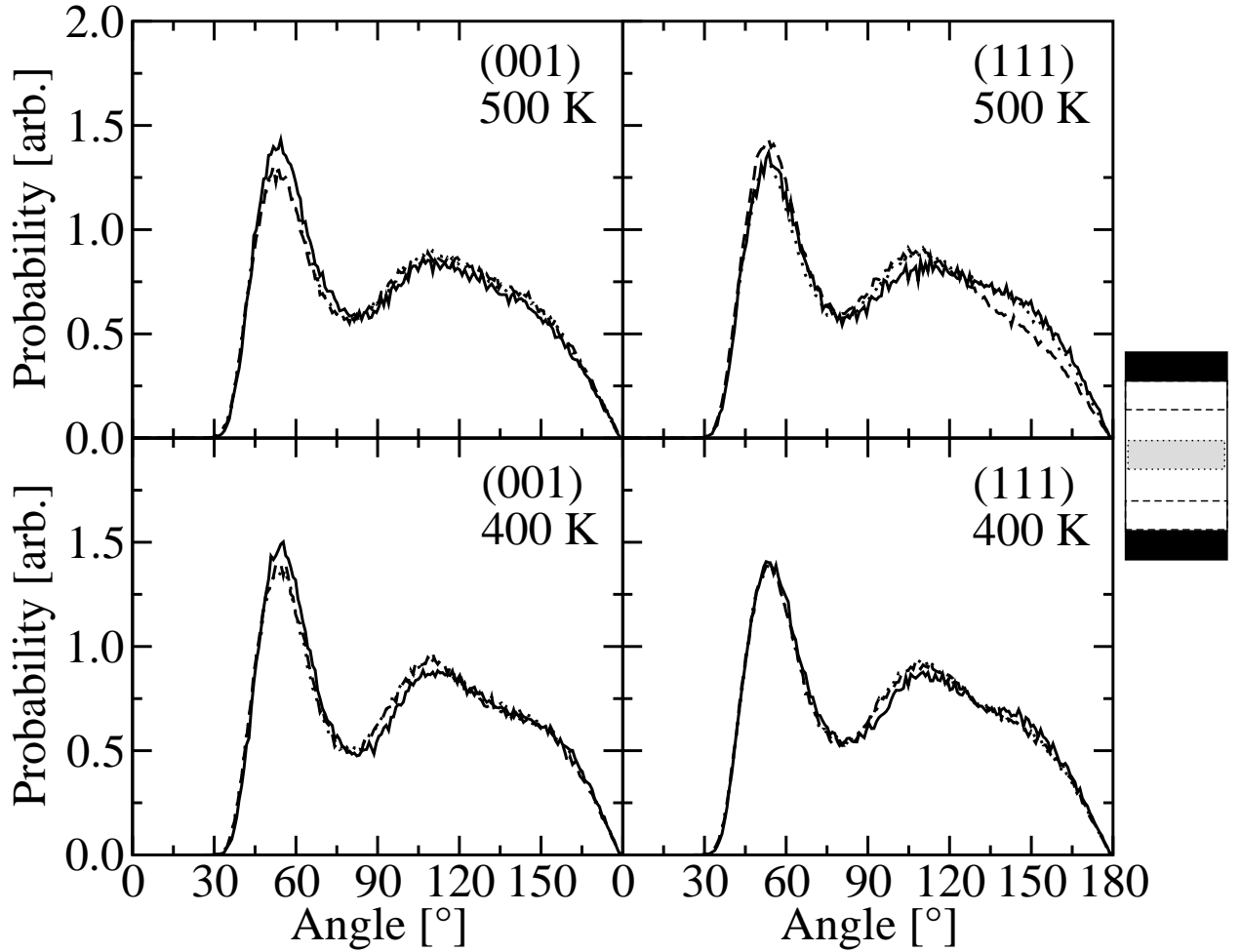}
\caption{\label{fig:bond_angle_distribution_slab} Bond angle
  distributions calculated in various slices of the \textit{ab initio}
  MD slabs. Solid curves: surface layers; dashed curves: 2nd surface
  layers; dotted curves: averages over the inner layers.}
\end{center}
\end{figure}

Bond angle distributions for the slabs appear in
Fig.~\ref{fig:bond_angle_distribution_slab}. There we compare the
distributions of angles for the surface layers, the 2nd surface
layers, and for layers in the interior of the slab. The distributions
are calculated using the same nearest neighbor sphere as considered in
determining the nearest neighbor distributions. To calculate the bond
angle distributions, the bond angles between an atom and each pair of
other atoms lying within the nearest neighbor sphere centered on the
first atom are determined. In general, comparing the surface layer,
2nd surface layer and inner bond angle distributions shows there to be
little difference between the bond angle distributions in the
different regions of the slabs at the same temperature. In comparing
the bond angle distributions at the different temperatures, the
separation of a small shoulder at large angles can be seen at the
lower temperature. This may signal an enhancement of the angles
($60^{\circ}$, $120^{\circ}$ and $180^{\circ}$), the three bond angles
characteristic of six-fold coordinated hexagonal ordering. However, it
should be noted that the bond angles that would appear in an
icosahedral fragment are $63.5^{\circ}$ and $116.5^{\circ}$, so it is
difficult to say whether the fairly small changes seen here in the
bond angle indicate strengthening of 5 or 6-fold ordering.

\begin{figure}[!hbt]
\begin{center}
\includegraphics[clip,width=0.9\columnwidth]{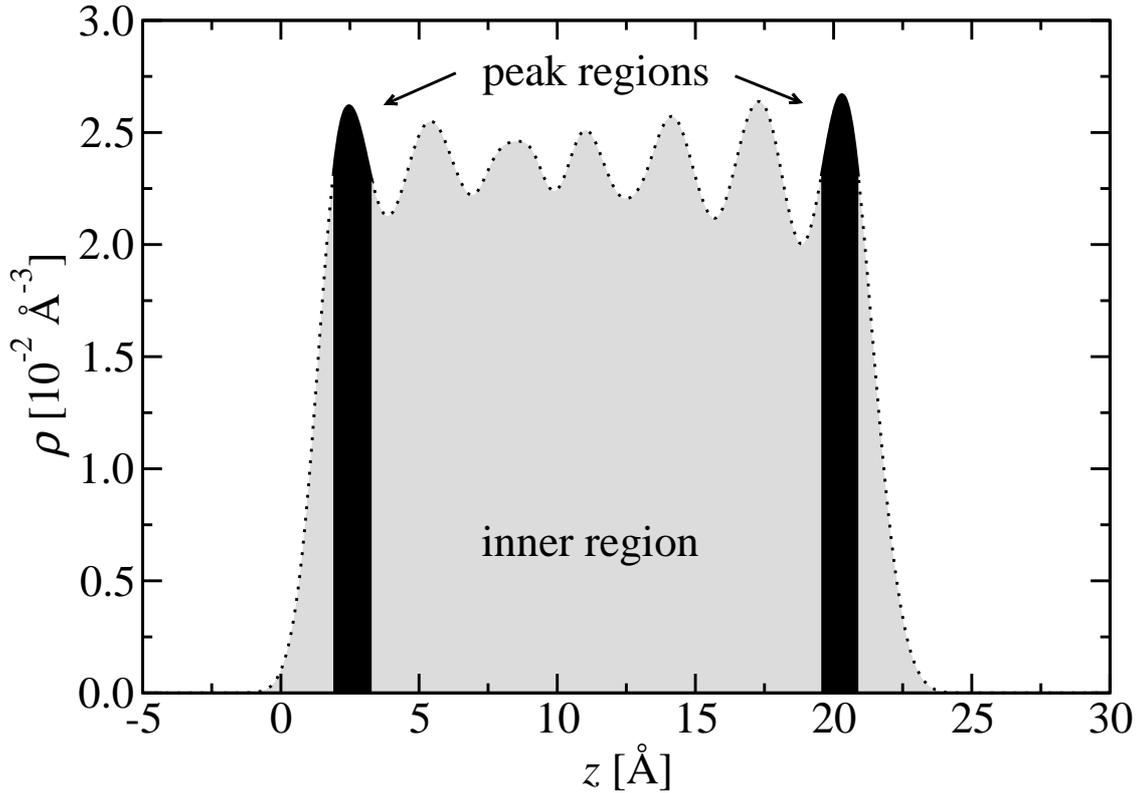}
\caption{\label{fig:interface_pic3} Illustration of the ``peak''
  regions (black).}
\end{center}
\end{figure}

\begin{figure}[!hbt]
\begin{center}
\includegraphics[clip,width=0.9\columnwidth]{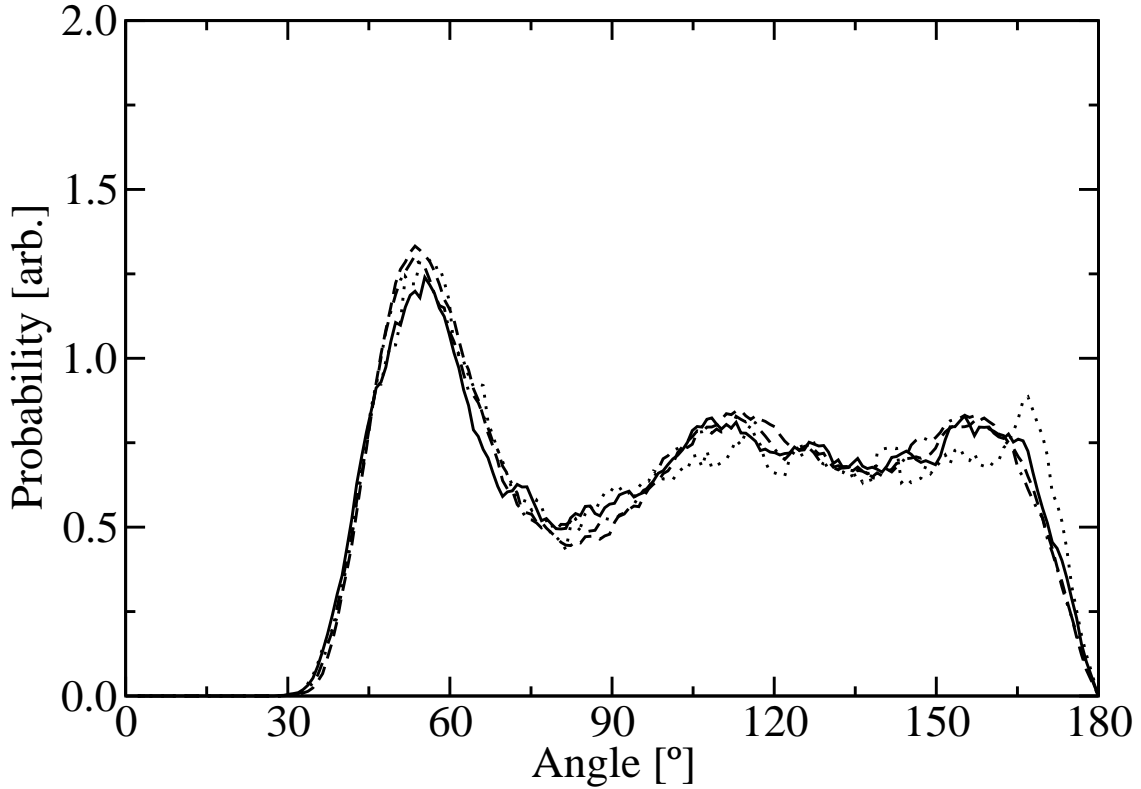}
\caption{\label{fig:angles_peaks_pic} Bond angle distributions
  calculated in the peak regions illustrated in
  Fig.~\ref{fig:interface_pic3} for \textit{ab initio} free surface
  simulations. The curves refer to the simulations as follows: solid:
  $(001)$, $T=500$~K; dotted: $(111)$, $T=500$~K; dashed: $(001)$,
  $T=400$~K; dot-dashed: $(111)$, $T=400$~K.}
\end{center}
\end{figure}

If attention is confined to small regions located right at the
outermost peaks (as illustrated in Fig.~\ref{fig:interface_pic3}), the
movement of concentration towards the angles $60^{\circ}$,
$120^{\circ}$ and $180^{\circ}$ becomes more pronounced, as shown in
Fig.~\ref{fig:angles_peaks_pic}.

When we restrict attention to the narrow regions centered on the
outermost (surface) peaks, the quality of the statistics is
correspondingly lowered. Here finite size effects in the statistics
are clearly much more important, as can be seen easily from the level
of noise in the bond angle distributions found in the ``peak''
regions.

\subsection{Classical MD liquid surface simulations \label{subsec:classical_free_surface}}

To complement the \textit{ab initio} simulation results, we also
perform explorations using classical MD simulations. Because the
statistics that can be collected with the (less accurate) classical
simulations are much longer than those obtainable with \textit{ab
  initio} MD, classical MD provides us with a convenient check on
simulation time effects in the statistics (also larger numbers of
atoms can be simulated with classical simulations).

The classical surface samples were obtained -- in analogy with the
\textit{ab initio} setup procedure -- by melting a crystal lattice at
high temperature, followed by addition of a vacuum region of thickness
$\sim$11 \r{A} along the $z$ direction to form a slab of atoms, and
equilibration at the temperatures of interest. The cells have the same
shape as the (001) cells used in the \textit{ab initio} simulations,
but are larger, containing 720 atoms (the atomic slabs are roughly
twice as thick as in the \textit{ab initio} simulations and are about
20 $\%$ larger in each of the transverse directions; the corresponding
solid slab would contain 20 layers parallel to the surfaces, and 12
layers in the $x$ and $y$ directions). The length of time simulated
was roughly 10 times that achieved in the \textit{ab initio} surface
simulations. The potential used to model the interactions between the
atoms in the classical simulations was that fitted for sodium by
Chac\'{o}n \textit{et al}
\cite{chacon1,chacon2,chacon3,na_paper_jpcmlett,na_paper_layers}.

\begin{figure}[!hbt]
\begin{center}
\includegraphics[clip,width=\columnwidth]{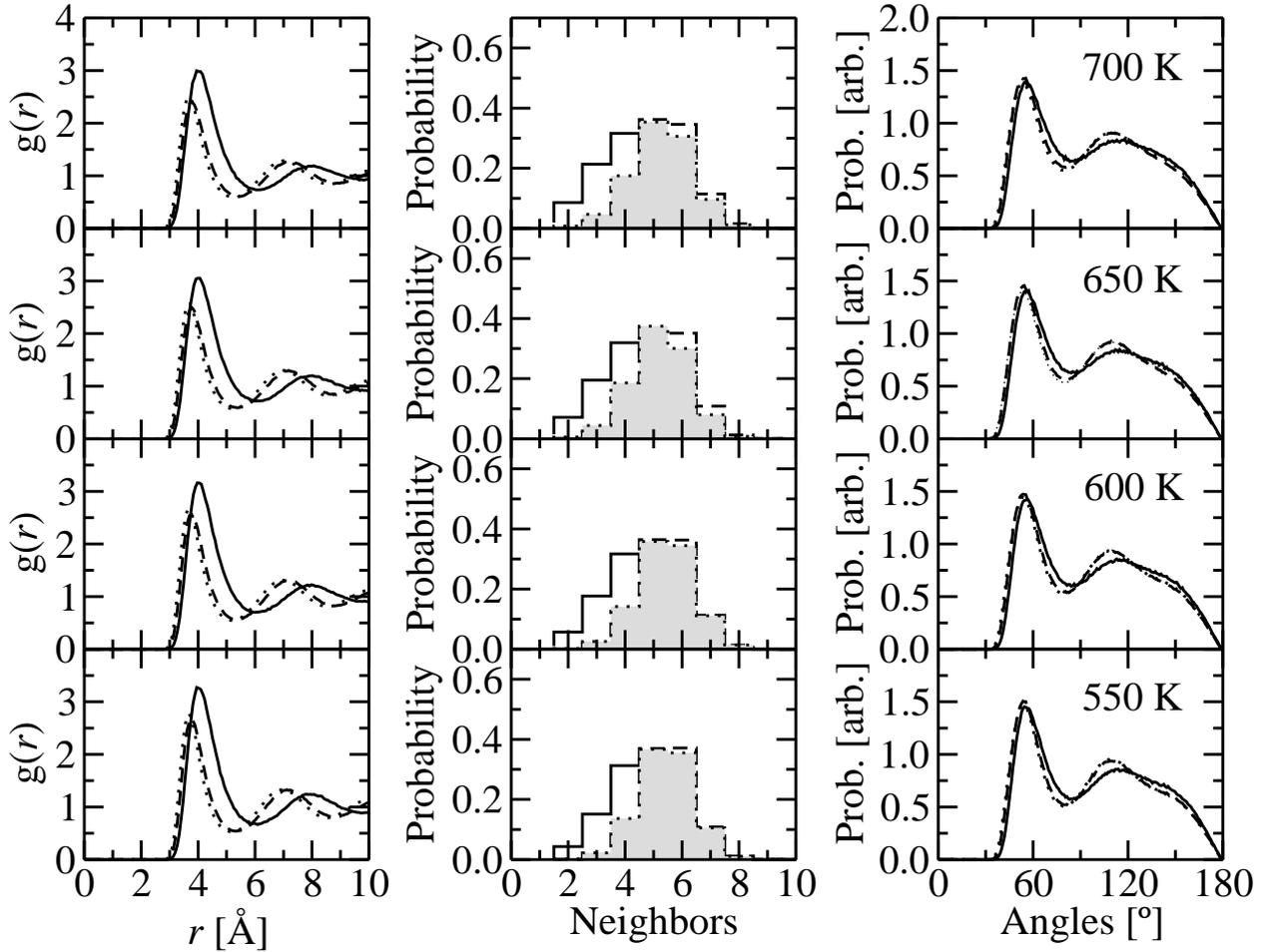}
\caption{\label{fig:class_angles_pic} Transverse pair correlation
  functions (left panels), nearest neighbor distributions (middle
  panels), and bond angle distributions (right panels) calculated for
  classical MD surface simulations. The temperatures are, from top to
  bottom: 700 K, 650 K, 600 K, and 550 K. In all cases, solid profiles
  refer to the surface layers, dashed profiles to the 2nd surface
  layers, and dotted profiles to averages over the inner layers.}
\end{center}
\end{figure}

In Fig.~\ref{fig:class_angles_pic} we plot the pair correlation
functions, nearest neighbor distributions, and bond angle
distributions for classical simulations at temperatures $T=550$ K,
$T=600$ K, $T=650$ K, $T=700$ K (the melting temperature for the pair
potential is approximately $T=650$ K). To obtain the averages over the
inner regions of the slabs we have used the discernible layers in
going from the surface into the slab; as the slabs in the classical
simulations are twice as thick as those in the \textit{ab initio}
simulations, there are regions in the centers where a layered
structure is not clear, that is, the bulk liquid limit has been
reached in those regions.

The pair correlation functions obtained from our classical surface
simulations show -- in addition to shifts in the positions of the
second peaks to larger radii -- shifts in the \textit{first} peaks to
larger radii in the outermost surface layer, with there being very
little difference between the profiles in the 2nd surface layers and
those in the inner layers. The shifts in the first peaks are
consistent with the behavior seen in the profiles for sodium shown in
Fig.~7 of Ref.~\onlinecite{gonzalez3}, though the shifts we see in the
first peak are larger than theirs; this is however different from the
behavior seen in our \textit{ab initio} free surface simulations
(cf.~Fig.~\ref{fig:corr_slab_pic}).

The nearest neighbor distributions are centered on 5, and most atoms
have 5 nearest neighbors, though there are significant fractions of
atoms having coordination numbers of 4 and 6. At the surface, the
nearest neighbor distributions are shifted towards lower
coordination. As the temperature is reduced, the distribution -- still
centered on 5 -- becomes slightly narrower, suggesting the degree of
order is increased. The average coordination numbers obtained by
averaging the nearest neighbor distributions are shown in Table
\ref{tab:nearest_neighbours_table_classical}.

\begin{table}[!htb]
\begin{center}
\begin{tabular}{cccc}
\hline $T$ [K] & Surface Layer & 2nd Surface Layer & Inner \\
\hline 550     &  4.4  &  5.5  &  5.4 \\
       600     &  4.3  &  5.5  &  5.4 \\
       650     &  4.2  &  5.4  &  5.2 \\
       700     &  4.1  &  5.4  &  5.2 \\
\hline
\end{tabular}
\end{center}
\caption{\label{tab:nearest_neighbours_table_classical} Coordination
  numbers $n_{c}$ obtained by averaging nearest neighbor distributions
  from the classical simulations shown in
  Fig.~\ref{fig:class_angles_pic}.}
\end{table}

The classical simulation coordination numbers are significantly
smaller for the surface layers than the second surface and inner
layers. Much smaller differences are seen between the coordination
numbers in the second surface layers and the inner layers. Notably,
the differences between the surface layer and the second surface and
inner layers in the classical simulations are greater than in the
\textit{ab initio} simulations, though comparable to the differences
between the surface, second surface and inner layers presented in
Table IV of Ref.~\onlinecite{gonzalez3}. This is reminiscent of the
density profiles normal to the surface \cite{na_paper_layers}, in that
the density profiles in the classical simulations were qualitatively
the same as those in Ref.~\onlinecite{gonzalez3} -- with the surface
peak significantly lower in height than the 2nd surface peak and the
inner peaks -- in contrast to the density profiles seen in our
\textit{ab initio} simulations, where the outermost peaks were at
roughly the same heights as the second surface and inner
peaks. Additionally, we see decreases in the coordination numbers in
the surface layers, the second surface layers, and the inner layers as
the temperature is increased in the classical simulations.

The bond angle distributions show the same sort of behavior as
observed in the \textit{ab initio} free surface simulations, that is,
the separation of shoulders in the distributions in the surface
layer. Here it is clear that the bond angle distributions in the 2nd
surface layer and the inner layers coincide, and there is a difference
between the surface layer and the second surface and inner layers. The
change in the bond angle distributions at the surface suggests more
convincingly a movement of weight towards the angles consistent with
6-fold coordination.

\subsection{\textit{Ab initio} MD liquid-solid interface simulations \label{subsec:liq_sol_interface}}

In addition to simulating the free liquid surface, we also consider MD
simulations of a model representing the solid-liquid interface,
constructed by placing a close-packed layer of fixed atoms within a
bulk cell. A random time step from a well-equilibrated $T=500$~K bulk
liquid \textit{ab initio} MD simulation is taken, and 24 of the atoms
are placed onto the plane $z=0$ in a distorted hexagonal 2D
arrangement (this is dictated by the use of a cell with a square
planar cross-section). Our initial motivation for constructing such a
model was that for sodium the electronic charge density will be almost
continuous normal to the ``interface'' (that is, there would be no
rapid decrease in the electronic density as there is at the free
liquid surface), allowing us to cleanly examine geometrical
confinement of the liquid atoms by the rigid atoms of the wall
\cite{na_paper_jpcmlett,na_paper_layers}. At this interface, layer
formation in the liquid part is seen; indeed layer formation is
significantly more pronounced in these solid-liquid interface
simulations at $T=400$~K than in the free surface simulations, and
there is considerable layer formation even at the high temperature of
800 K.

\begin{figure}[!hbt]
\begin{center}
\includegraphics[clip,width=\columnwidth]{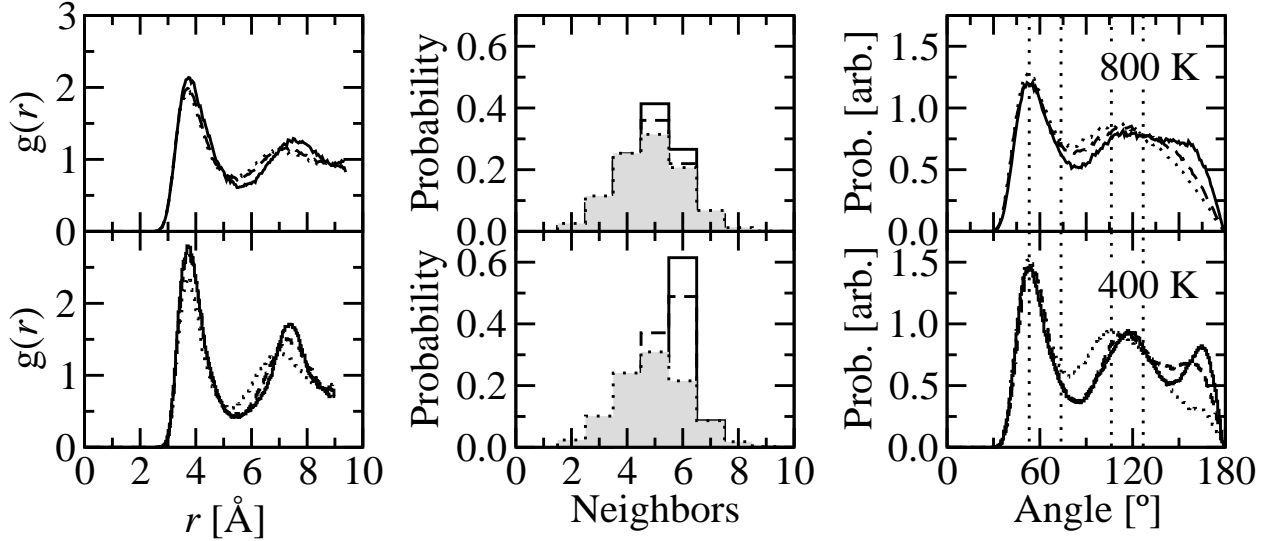}
\caption{\label{fig:angles_fixed_layers_pic} Pair correlation
  functions (left panels), nearest neighbor distributions (center
  panels) and bond angle distributions (right panels) for solid-liquid
  interface model. The upper panels refer to $T=800$~K, the lower
  panels to $T=400$~K. The solid curves refer to the layers of liquid
  atoms nearest to the wall, the dashed curves to averages over the
  inner layers lying between \textit{minima} in the density profiles,
  the dotted curves to averages over the inner layers lying between
  \textit{maxima} in the density profiles. The vertical dotted lines
  in the bond angle distribution plots indicate the bond angles within
  the distorted hexagonal fixed layer.}
\end{center}
\end{figure}

It is of course also interesting to consider the arrangement of the
liquid atoms parallel to the wall, in particular to see if it is
influenced by the fixed 2D order of the wall. In
Fig.~\ref{fig:angles_fixed_layers_pic} the pair correlation functions,
and bond angle and nearest neighbor distributions for our model
solid-liquid interface are shown. For the two simulations, we present
pair correlation functions, and bond angle and nearest neighbor
distributions for the layer of liquid atoms closest to the wall, and
averaged over inner liquid layers defined as lying within (i).~the
maximum-maximum regions, and (ii).~the minimum-minimum regions.

The bond angle distributions for the solid-liquid interface show
noticeable differences between the liquid layer closest to the fixed
layer and those in the inner region. The quasi-hexagonal (that is,
slightly distorted) arrangement of the atoms in the fixed layer means
the bond angles in it are not exactly those of the perfect hexagonal
arrangement ($60^{\circ}$, $120^{\circ}$ and $180^{\circ}$), but
instead: $53.1^{\circ}$, $73.7^{\circ}$, $106.3^{\circ}$,
$126.9^{\circ}$ and $180^{\circ}$. In the liquid layers nearest the
fixed layer, the bond angle distributions indicate a favoring of
angles near $120^{\circ}$ and an increase in weight at angles toward
$180^{\circ}$; the same type of effect of which we saw some hint in
the free surface simulations. The shift in weight towards higher
angles is enhanced as the temperature is lowered. This indicates that
the 2D order in the liquid layers adjacent to the fixed layer is
influenced by the structure of that layer: upon moving further from
the fixed layer into the liquid, the degree of order is reduced,
meaning the 2D order becomes less definite. Also as the temperature is
lowered, the disorder in the liquid part is lowered, meaning the
influence of the 2D order of the fixed layer is greater.

The average coordination numbers in the various liquid layers obtained
by averaging the nearest neighbor distributions are shown in Table
\ref{tab:nearest_neighbours_table_fixed_layer}. In the liquid layer
adjacent to the fixed layer, the coordination is close to the value of
six that the atoms within the fixed layer have, especially at
$T=400$~K; it decreases in the inner layers, and also when the
temperature is increased to 800 K. The coordination numbers in the
minimum-minimum layers are larger than in the maximum-maximum layers,
which is consistent with the idea of the atoms being ordered
\textit{within} the layers defined by the density peaks, where there
are more atoms.

\begin{table}[!htb]
\begin{center}
\begin{tabular}{cccc}
\hline $T$ [K] & First liquid layer & Inner: min-min & Inner: max-max \\
\hline 400     &  5.8  &  5.5  &  4.9 \\
       800     &  5.0  &  4.9  &  4.8 \\
\hline
\end{tabular}
\end{center}
\caption{\label{tab:nearest_neighbours_table_fixed_layer} Coordination
  numbers $n_{c}$ obtained by averaging the nearest neighbor
  distributions from the classical simulations shown in
  Fig.~\ref{fig:angles_fixed_layers_pic}.}
\end{table}

The pair correlation functions for the fixed layer simulations shown
in Fig.~\ref{fig:angles_fixed_layers_pic} display small differences in
the position of the first peak between the liquid layer adjacent to
the fixed layer and the inner layers, but there is a definite shift in
the position of the second peak to larger values in the first liquid
layer compared to the inner layers. This is the same type of behavior
we saw in comparing the pair correlation functions for the surface
layers and the inner layers in the density functional theory free
surface simulations. At $T=400$~K, the first minimum (to the right of
the first peak) is quite deep, and the first peak is higher in the
liquid layer adjacent to the wall, which supports the idea that the 2D
order in the fixed layer influences the liquid layer adjacent to it
(this deepening of the minimum and increase in the height of the
maximum is a signature of the correlation function in the fixed layer,
that consists of a series of spikes at the nearest neighbor, second
nearest neighbor, etc.~ distances): what we see in the liquid layer
adjacent to the fixed layer is a smeared version of this.

Now we attempt to rationalize the 2D ordering behavior suggested by
our simulations. The nearest neighbor and bond angle distributions
given above, suggesting predominantly 5-fold and 6-fold coordinated
atomic arrangements, can be explained by considering an ordering
reminiscent of the hexatic phase to be present at the surface.

\begin{figure}[!bt]
\begin{center}
\includegraphics[clip,width=0.9\columnwidth]{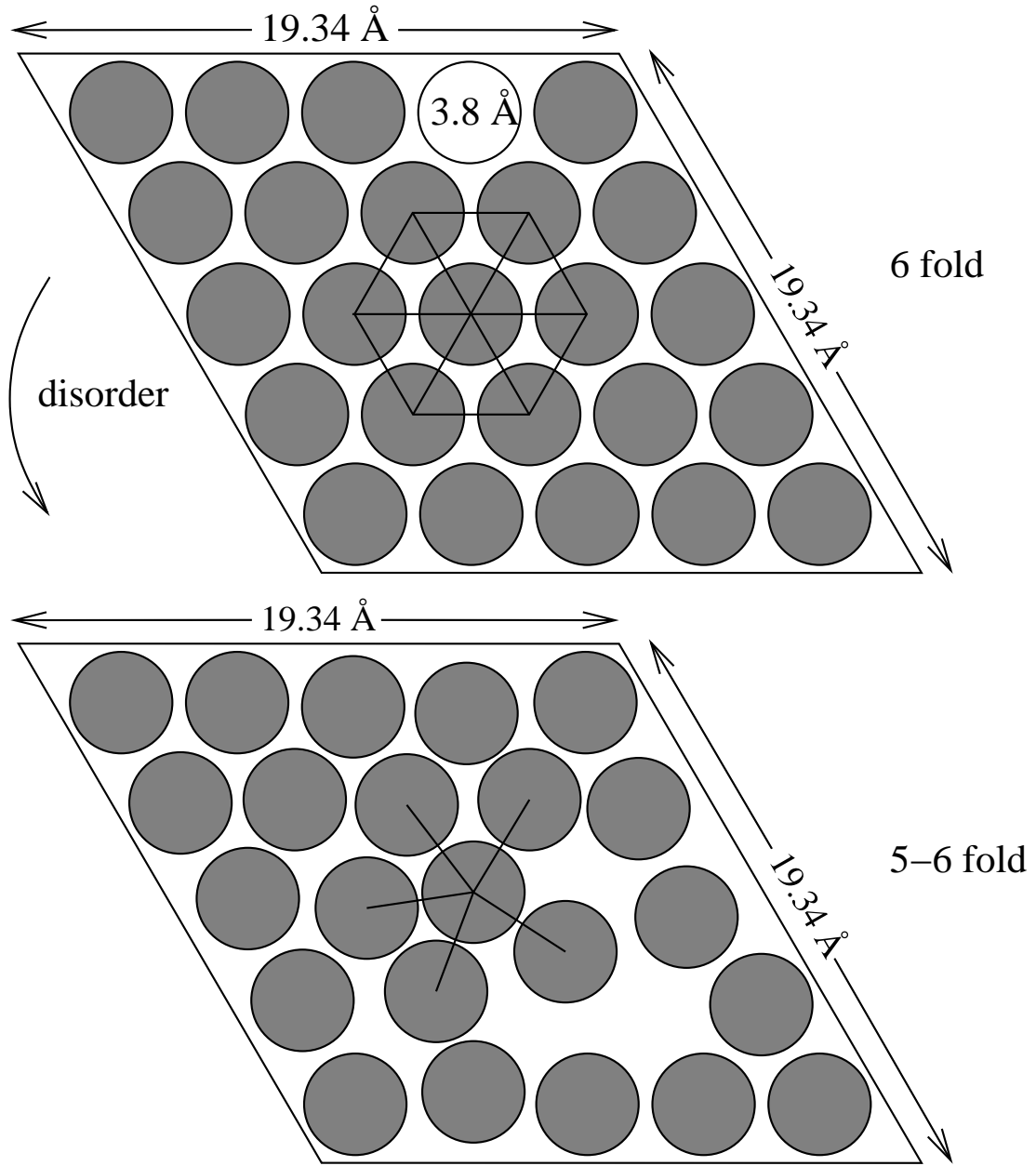}
\caption{\label{fig:disordered_hex_pack} Schematic illustration of
  taking a hexagonally close-packed set of atoms (upper part),
  removing a couple of them, and introducing some disorder into the
  positions of those remaining, to obtain a distorted hexagonal
  ordering (lower part). The cell is of the $(111)$ shape and the cell
  dimensions are those appropriate for $T=500$~K.}
\end{center}
\end{figure}

Firstly we take the most probable spacing between atoms in the surface
layer indicated by the pair correlation functions, $3.8$~\r{A}. We
consider a close-packed hexagonal arrangement of atoms with this
spacing (this will contain 25 atoms for perfect hexagonal
arrangement), and then remove a couple of these atoms and introduce
some disorder into the arrangement (as indicated schematically in
Fig.~\ref{fig:disordered_hex_pack}). This cartoon is intended to give
a qualitative indication of what the atomic arrangement might look
like; the nearest neighbor distance is used as a guide to the size of
the atoms.  In this way, we estimate that the number of atoms in such
a layer should be $\sim$$24-25$
\cite{na_paper_jpcmlett,na_paper_layers}, and the average coordination
number would be a bit less than 6.

For slabs consisting of around 160 atoms formed into
quasi-close-packed layers, a total of 7 layers would be expected. This
is in agreement with the numbers of layers present in the slab
simulations as indicated by the density profiles
\cite{na_paper_layers}.

Fig.~5 of Ref.~\onlinecite{jpcm_review} shows 2D density plots
calculated at the surfaces -- that is, at the positions of the surface
peaks in the density profiles -- for the (001) \textit{ab initio} slab
simulations discussed in this paper. Inspection of those plots shows
that there is considerable structure in the surface layers, with
regions consistent with five- and six-fold coordinated ordering, in
accordance with the bond angle and nearest neighbor distributions
presented above, and our schematic picture of distorted hexagonal
ordering in the surface layer.

\section{Conclusions\label{sec:conclusions_section}}

The data we have presented in this paper suggest a weak tendency
toward two-dimensional atomic ordering within the layers formed at
liquid metal surfaces, especially the surface layer. Nearest neighbor
distributions indicate predominantly a mixture 5-fold and 6-fold
coordinated sites at the surface (though there are notable fractions
of 4-fold coordinated sites), with the average coordination within the
surface layers being just over 5. The fraction of 4-fold coordinated
sites decreased as the temperature is lowered, while the fractions of
5- and 6-fold coordinated sites increase at the surface. Bond angle
distributions indicate increased tendency toward the three bond angles
of a hexagonally close-packed two-dimensional structure, at the
surface. We have rationalized this data from the \textit{ab initio} MD
simulations by considering a slight disordering of a
hexagonally-ordered two-dimensional structure.

The classical and \textit{ab initio} DFT free liquid surface
simulations we have described show broadly similar results in that
both types of simulation produce a layered surface density, and are
consistent with with the results of orbital-free density functional
theory simulations \cite{gonzalez,gonzalez2,gonzalez3}. There are
however differences between the \textit{ab initio} free surface
density profiles and the classical and orbital-free DFT results in
that the \textit{ab initio} densities show outermost peaks larger than
or of comparable height to the subsequent peaks away from the surface,
whereas the classical and orbital-free simulations consistently show
surface peaks of lower height than the subsequent peaks. The analysis
of the experimental results
\cite{expt1,expt2,expt3,expt4,expt5,expt6,expt7,expt8,expt9,expt10,expt11}
gives density profiles with surface peaks with heights greater than
those of the subsequent peaks.

\begin{acknowledgments}
BGW thanks the New Zealand Foundation for Research Science and
Technology and the Cambridge Overseas Trust for studentships. NM
acknolwedges computational support from NSF DMR-0414849 and PNNL
EMSL-UP-9597.
\end{acknowledgments}

\bibliographystyle{plain}
\bibliography{na_paper_inplane}

\end{document}